# Superconductivity of F-substituted $Ln$OBiS$_2$ ($Ln$ = La, Ce, Pr, Nd, Yb) compounds

D. Yazici, K. Huang, B. D. White, A. H. Chang, A. J. Friedman, and M. B. Maple

*Department of Physics, University of California, San Diego, La Jolla, CA 92093*

E-mail: mbmaple@physics.ucsd.edu

Polycrystalline samples of F-substituted $Ln$OBiS$_2$ ($Ln$ = La, Ce, Pr, Nd, Yb) compounds were synthesized by solid-state reaction. The samples were characterized by x-ray diffraction measurements and found to have the ZrCuSiAs crystal structure. Electrical resistivity and magnetic susceptibility measurements were performed on all of the samples and specific heat measurements were made on those with $Ln$ = La, Ce, and Yb. All of these compounds exhibit superconductivity in the range 1.9 K – 5.4 K, which has not previously been reported for the compounds based on Ce, Pr, and Yb. The YbO$_{0.5}$F$_{0.5}$BiS$_2$ compounds was also found to exhibit magnetic order at ~2.7 K that apparently coexists with superconductivity below 5.4 K.

## 1. Introduction

Soon after the discovery of superconductivity in the BiS$_2$-based compound Bi$_4$O$_4$S$_3$ with a superconducting critical temperature ($T_c$) of 8.6 K by Mizuguchi *et al*. [1], a new BiS$_2$-based superconductor, LaO$_{1-x}$F$_x$BiS$_2$, with the relatively high $T_c$ of 10.6 K was reported [2]. Recently, another BiS$_2$-based superconductor, NdO$_{1-x}$F$_x$BiS$_2$ (x = 0.1 - 0.7) was found to have a maximum $T_c$ of 5.6 K at x ≈ 0.3 [3]. These materials have a layered crystal structure composed of superconducting $BiS_2$ layers and blocking layers of $Bi_4O_4(SO_4)_{1-x}$ for $Bi_4O_4S_3$ and $LnO$ for $LnO_{1-x}F_xBiS_2$ ($Ln$ = La, Nd). This is similar to the situation encountered in the high-$T_c$ layered cuprate and Fe-pnictide superconductors, in which the superconductivity primarily resides in $CuO_2$ planes and Fe-pnictide layers, respectively.

Polycrystalline samples of F-substituted $LnOBiS_2$ ($Ln$ = La, Ce, Pr, Nd, Yb) compounds were synthesized by solid-state reaction. The compounds were characterized by x-ray diffraction measurements and found to have the ZrCuSiAs crystal structure. Electrical resistivity and magnetic susceptibility measurements were made on all of the samples, while specific heat measurements were made on the samples based on La, Ce and Yb. All of these compounds exhibit superconductivity in the range 1.9 K – 5.4 K, while the compounds with Ce, Pr, and Yb are new superconductors. The $YbO_{0.5}F_{0.5}BiS_2$ sample was also found to exhibit magnetic order at ~2.7 K that apparently coexists with superconductivity below 5.4 K.

## 2. Experimental Details

Polycrystalline samples of $LnO_{1-x}F_xBiS_2$ ($Ln$ = La, Ce, Pr, Nd, Yb) with x = 0.5 were prepared by solid-state reaction using powders of $Ln_2O_3$ ($Ln$ = La, Pr, Nd, Yb) (99.9%), $CeO_2$ (99.9%) for $CeO_{1-x}F_xBiS_2$, $LnF_3$ (99.9%), $Ln_2S_3$ (99.9%), and $Bi_2S_3$ (99.9%). The $Bi_2S_3$ precursor powders were prepared by reacting Bi (99.99%) and S (99.9%) together at 500 °C in an evacuated quartz tube for 10 hours. The four $Ln_2S_3$ precursor powders were prepared by reacting $Ln$ chunks and S grains at 800 °C in evacuated quartz tubes for 12 hours. The starting materials with nominal composition of $LnO_{0.5}F_{0.5}BiS_2$ ($Ln$ = La,

Ce, Pr, Nd, Yb) were weighed, thoroughly mixed, pressed into pellets, sealed in evacuated quartz tubes, and annealed at 800 °C for 10 hours. The products were then ground, mixed for homogenization, pressed into pellets, and annealed again in evacuated quartz tubes at 800 °C for 10 hours. X-ray powder diffraction measurements were made using an x-ray diffractometer with a Cu $K_\alpha$ source to assess phase purity and to determine the crystal structure of the $LnO_{0.5}F_{0.5}BiS_2$ ($Ln$ = La, Ce, Pr, Nd, Yb) compounds. Electrical resistivity $\rho(T)$ measurements were performed using a four-wire technique in a liquid $^4$He dewar in the temperature range 1.1 K $\leq T \leq$ 300 K. The temperature dependence of the magnetization was measured using a Quantum Design (QD) Magnetic Property Measurement System (MPMS) with an applied magnetic field of 5 Oe. Specific heat measurements were made using a standard thermal relaxation technique in the temperature range 1.8 K $\leq T \leq$ 50 K in a QD Dynacool Physical Property Measurement System (PPMS).

## 3. Results and discussion

Figure 1 displays a representative powder x-ray diffraction (XRD) pattern for $LaO_{0.5}F_{0.5}BiS_2$. The broad feature in the XRD background between 25 and 35 degrees is associated with the glass substrate. Although the XRD pattern contains a small number of impurity peaks belonging to $LaF_3$, which are indicated by arrows in Fig. 1, all of the major peaks can be indexed to the space group *P*4/*nmm*. Similar patterns (not shown) were obtained for the other compounds with similar impurity levels of $LnF_3$. The lattice constants were estimated from 2Θ values of selected Bragg reflections and are listed in Table 1. The crystal structure of $LaOBiS_2$ is displayed in the inset of Fig. 1.

Electrical resistivity $\rho(T)$ data for the $LnO_{0.5}F_{0.5}BiS_2$ ($Ln$ = La, Ce, Pr, Nd, Yb) samples are shown in Fig. 2. For the samples with $Ln$ = La, Ce, and Yb, $\rho$ increases with decreasing temperature until the onset of the superconducting transition at temperatures of 3.6, 2.2, and 5.4 K for $Ln$ = La, Ce, and Yb, respectively. For the other samples with $Ln$ = Pr and Nd, $\rho$ first decreases slightly with decreasing temperature down to ~230 K and then increases with decreasing temperature until the onset of the superconducting transition at temperatures of 4.5 and 4.4 K for $Ln$ = Pr and Nd, respectively. The value of $T_c$ was defined as the temperatures where $\rho$ drops to 50% of its normal state value. These values are tabulated in Table 1. Measurements of electrical resistivity (not shown) were also performed down to 1.8 K in applied magnetic fields for samples of $LaO_{0.5}F_{0.5}BiS_2$ and $YbO_{0.5}F_{0.5}BiS_2$. The orbital critical fields for these compounds were inferred from their $T_c$'s and initial slopes of $H_{c2}$ with respect to $T$, such that values of 0.757 T and 1.38 T, respectively, were calculated using the Werthamer-Helfand-Hohenberg theory [4].

To determine whether the superconductivity is a bulk phenomenon, zero field cooled (ZFC) and field-cooled (FC) measurements of the magnetic susceptibility $\chi(T)$ were made in a field of 5 Oe for the $LnO_{0.5}F_{0.5}BiS_2$ ($Ln$ = La, Ce, Pr, Nd, Yb) samples, as shown in Figs. 3 (La, Ce, Pr, Nd) and 4 (Yb). For the $LnO_{0.5}F_{0.5}BiS_2$ ($Ln$ = La, Ce, Pr, Nd) samples (see Fig. 3), zero field cooled (ZFC) measurements yield diamagnetic screening signals with $T_c$ onset values that are consistent with the $\rho(T)$ data, while field cooled (FC) measurements reveal hardly any change in $\chi(T)$ in the superconducting state relative to the normal state, indicative of strong vortex pinning. In contrast, the ZFC and FC $\chi(T)$ data for $YbO_{0.5}F_{0.5}BiS_2$ both reveal changes and hysteresis at $T_c$ and peaks at lower temperature ~2.7 K that appear to be due to magnetic ordering. This interpretation is further supported by a sharp feature in the specific heat near ~2.7 K and shown in Fig. 7. That magnetic ordering is probably antiferromagnetic in nature, since the superconductivity persists to temperatures below ~2.7 K and appears to coexist with the magnetic order. For superconductors containing ions that order ferromagnetically, such as $ErRh_4B_4$ and $HoMo_6S_8$, the ferromagnetism destroys the superconductivity at a second critical temperature $T_{c2}$ of the order of the Curie temperature that is lower than the critical temperature $T_{c1}$ at which the sample first became superconducting [5,6].

Specific heat $C(T)$ data for $LnO_{0.5}F_{0.5}BiS_2$ samples with $Ln$ = La, Ce, and Yb are displayed in Figs. 5, 6, and 7, respectively. $C(T)$ data for $LaO_{0.5}F_{0.5}BiS_2$, covering the temperature range 1.8 K to 50 K, are shown in Fig. 5. A fit of the expression $C(T) = \gamma T + \beta T^3$, where $\gamma$ is the electronic specific heat coefficient and $\beta$ is the coefficient of the lattice contribution, to $C/T$ data in the normal state, plotted as a function of $T^2$, is shown in the inset in the lower right hand side of Fig. 5. From the best fit, which is explicitly indicated by a line in the inset, we obtain values of $\gamma$ = 2.53 mJ/mol-$K^2$ and $\beta$ = 0.904 mJ/mol-$K^4$; the value of $\beta$ corresponds to a Debye temperature of $\Theta_D$ = 221 K. In the inset in the upper left hand side of the figure, the electronic contribution to the specific heat $C_e(T) = C(T) - \beta T^3$ is shown, which has been derived by subtracting the lattice contribution $\beta T^3$ from $C(T)$. There is a clear jump at $T_c$ = 2.93 K, determined from the idealized entropy conserving construction shown in the figure. This value of $T_c$ is close to the temperature at which the electrical resistivity of $LaO_{0.5}F_{0.5}BiS_2$ vanishes (see Fig. 2). The presence of the jump clearly indicates the bulk nature of superconductivity in this compound. The ratio of the jump to the electronic contribution to specific heat at $T_c$, $\Delta C/\gamma T_c$ = 0.94, was calculated using the jump in $C_e/T$ of 2.37 mJ/mol-$K^2$ seen in the inset of Fig. 5. This value for $\Delta C/\gamma T_c$ is less than the value of 1.43 predicted by the BCS theory.

The $C(T)$ data for $CeO_{0.5}F_{0.5}BiS_2$ are displayed between 1.8 K and 50 K in Fig. 6. There is an upturn at low temperature that may indicate the presence of incipient magnetic order. However, measurements must be made at temperatures below 1.8 K to unambiguously identify the nature of this anomaly. The upper inset of Fig. 6 highlights the behavior of $C/T$ as a function of $T$. Absent from these data is any evidence for a jump at the $T_c$'s obtained from either the electrical resistivity ($T_c$ = 1.9 K) or magnetic susceptibility ($T_c$ = 5(1) K) measurements. However, the jump in $C/T$ for $LaO_{0.5}F_{0.5}BiS_2$ was closest to the temperature where its electrical resistivity dropped completely to zero, and according to Fig. 2, this occurs in $CeO_{0.5}F_{0.5}BiS_2$ below our base temperature. The lower inset of Fig. 6 highlights the linear fit to $C/T$ vs. $T^2$, from which we extracted $\gamma$ = 58.1 mJ/mol-K$^2$ and calculated $\Theta_D$ = 224 K.

Specific heat $C(T)$ data for $YbO_{0.5}F_{0.5}BiS_2$ between 1.8 K and 50 K are shown in Fig. 7. In the inset in the lower right hand side of the figure, a linear fit to the $C/T$ vs $T^2$ plot yields values of $\gamma$ = 30.1 mJ/mol-K$^2$ and $\Theta_D$ = 186 K. Highlighted in the inset of Fig. 7 on the upper left hand side of the figure are $C(T)$ data, which exhibit a lambda-like anomaly near 2.65 K that appears to be due to antiferromagnetic order. The entropy $S_m(T)$ associated with the magnetic ordering was estimated by integrating $C_e(T)/T$ from 0 K to 10 K, where the electronic contribution $C_e(T)$ was derived from $C(T)$ by subtracting the phonon contribution $\beta T^3$. The contribution to $S_e(T)$ between 0 K and 1.8 K, the low temperature limit of the $C(T)$ measurements, was estimated by linear extrapolation of the $C_e(T)/T$ curve to 0 K and integrating from 0 K to 1.8 K. The resultant entropy $S_m(T)$ (not shown) was found to saturate to approximately 4.8 J/mol-K at ~10K which is 84 % of Rln2 associated with a Kramers doublet ground state. This large value of the entropy reveals that the magnetic ordering is intrinsic and not due to an impurity phase such as $Yb_2O_3$, YbOF or $Yb_2O_2S$ which exhibit antiferromagnetic order with Néel temperatures in the range $T_N$ = 2.3 - 2.5 K. No jump due to the superconducting transition was evident in the $C(T)$ data at the $T_c$ of ~ 5.4 K as determined from $\rho(T)$ and $\chi(T)$ measurements (see Figs. 2 and 4). This is probably a consequence of the superconducting transition being spread out in temperature due to sample inhomogeneity.

## 4. Summary

Polycrystalline samples of $LnO_{0.5}F_{0.5}BiS_2$ ($Ln$ = La, Ce, Pr, Nd, Yb) compounds were synthesized by solid-state reaction and found to have the ZrCuSiAs crystal structure. Electrical resistivity and magnetic susceptibility measurements were performed on all of the samples and specific heat measurements were made on those with $Ln$ = La, Ce, and Yb. All of these compounds exhibit superconductivity in the range 1.9 K – 5.4 K, which has not previously been reported for the compounds based on Ce, Pr, and Yb. The $YbO_{0.5}F_{0.5}BiS_2$ sample was also found to exhibit magnetic order (probably antiferromagnetic order) at ~2.7 K that apparently coexists with superconductivity below 5.4 K.

**Acknowledgements**

The authors gratefully acknowledge the support of the US Air Force Office of Scientific Research – Multidisciplinary University Research Initiative under Grant No. FA9550-09-1-0603 (funding superconductivity searches) and the US Department of Energy under Grant No. DE-FG02-04-ER46105 (funding materials synthesis and characterization).

| Compound | a (Å) | c (Å) | V (Å$^3$) | $T_c$ (K) Resistivity | $T_c$ (K) Susceptibility |
|---|---|---|---|---|---|
| LaO$_{0.5}$F$_{0.5}$BiS$_2$ | 4.0613 | 13.3157 | 219.6341 | 3.10 | 2.8(3) |
| CeO$_{0.5}$F$_{0.5}$BiS$_2$ | 4.0398 | 13.4513 | 219.5269 | 1.89 | 5(1) |
| PrO$_{0.5}$F$_{0.5}$BiS$_2$ | 4.0192 | 13.4238 | 216.8491 | 4.29 | 3.5(4) |
| NdO$_{0.5}$F$_{0.5}$BiS$_2$ | 4.0102 | 13.4468 | 216.2468 | 4.37 | 4.0(3) |
| YbO$_{0.5}$F$_{0.5}$BiS$_2$ | 3.9851 | 13.5057 | 214.4820 | 5.30 | 5.0(3) |

Table 1.

Lattice parameters *a* and *c*, unit cell volume *V*, and superconducting critical temperature $T_c$ of *Ln*O$_{0.5}$F$_{0.5}$BiS$_2$ compounds with *Ln* = La, Ce, Pr, Nd, and Yb. The values of $T_c$ are estimated from the temperatures where the electrical resistivity drops to 50% of its normal state value and the ZFC and FC data separate in magnetic susceptibility measurements.

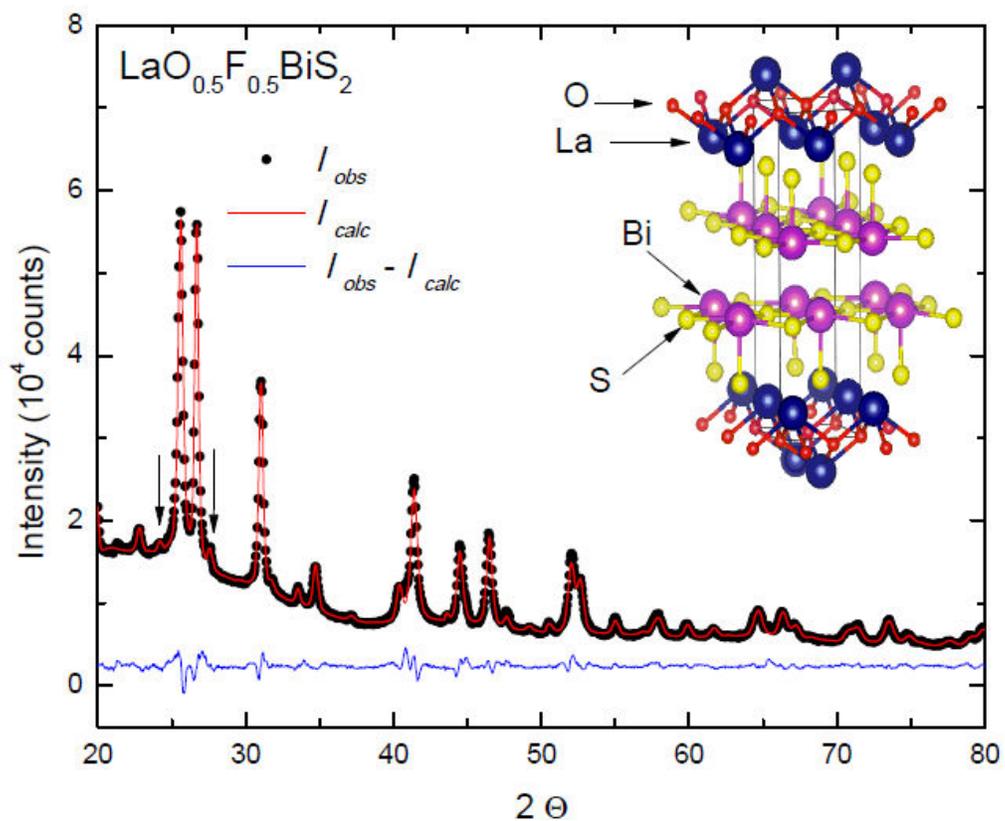

Figure 1. X-ray diffraction pattern for LaO$_{0.5}$F$_{0.5}$BiS$_2$ at room temperature. The black points indicate the observed intensity $I_{obs}$, the red line represents the calculated intensity $I_{calc}$, and the blue line indicates the difference $I_{obs} - I_{calc}$. The crystal structure is shown in the upper right hand part of the figure.

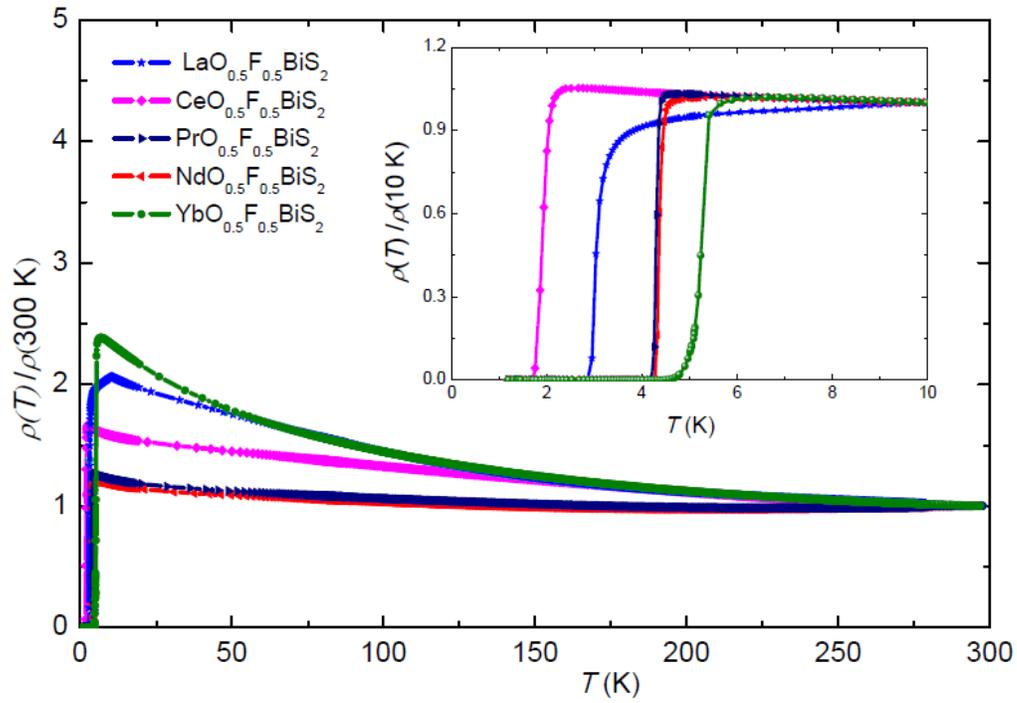

Figure 2. Electrical resistivity $\rho(T)$, normalized to its room temperature value $\rho(300\ K)$, vs temperature $T$ for $LnO_{0.5}F_{0.5}BiS_2$ ($Ln$ = La, Ce, Pr, Nd, Yb). The resistive superconducting transition curves are shown in the inset

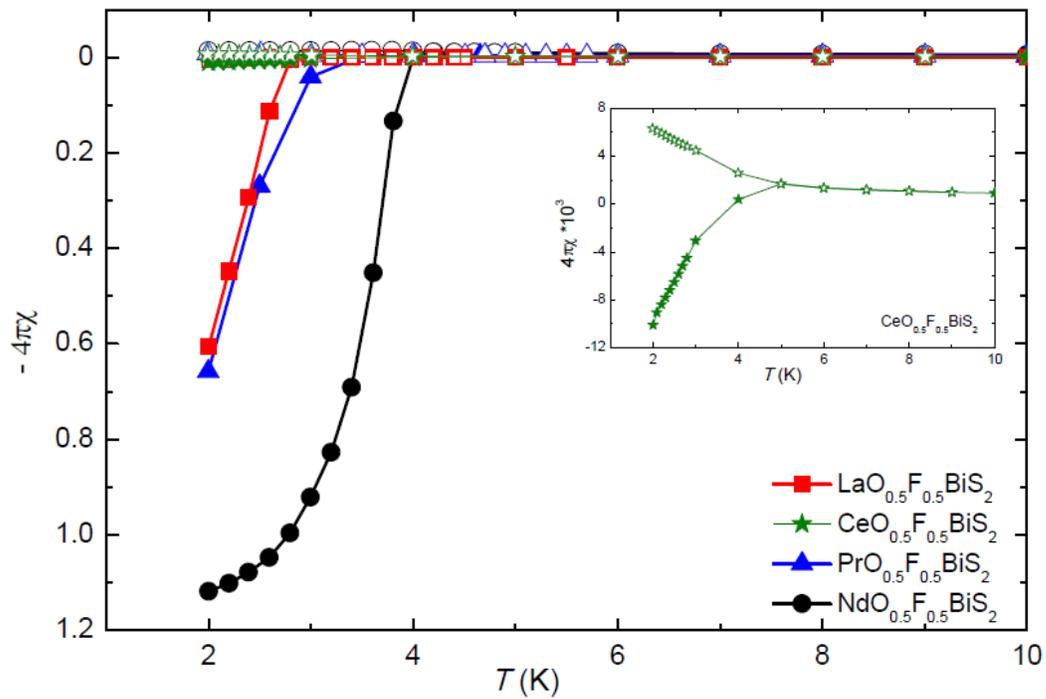

Figure 3. Magnetic susceptibility $\chi$ vs temperature $T$ for $LnO_{0.5}F_{0.5}BiS_2$ ($Ln$ = La, Pr, Nd), measured in the field cooled (FC) and zero field cooled (ZFC) conditions. The $\chi$ vs $T$ data for $CeO_{0.5}F_{0.5}BiS_2$ are shown in the inset.

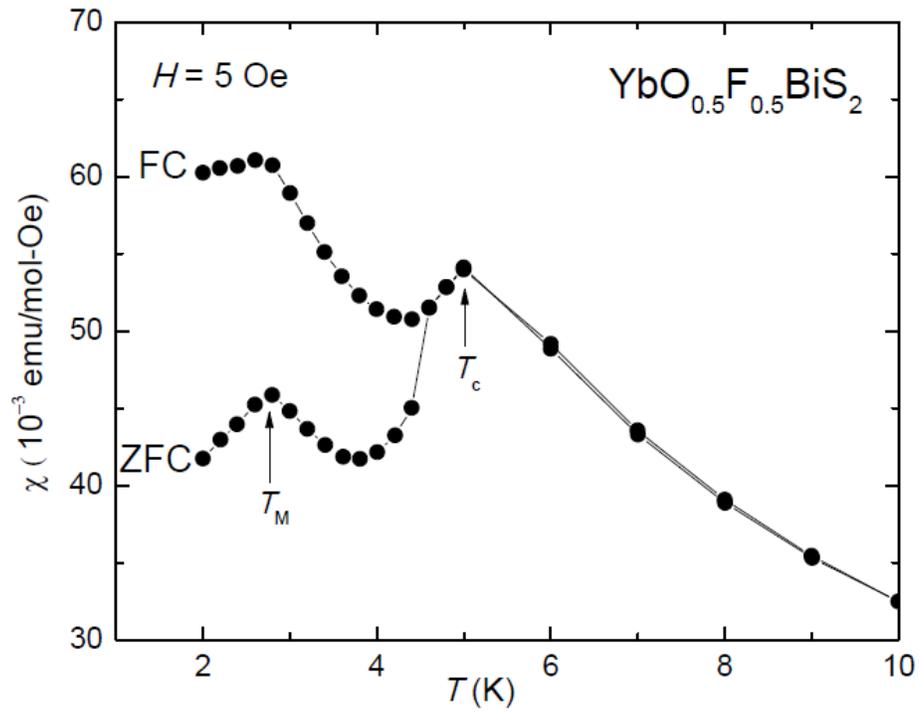

Figure 4. Magnetic susceptibility $\chi$ vs temperature $T$ for $YbO_{0.5}F_{0.5}BiS_2$, measured in the field cooled (FC) and zero field cooled (ZFC) conditions. The superconducting critical temperature $T_c$ is indicated in the figure.

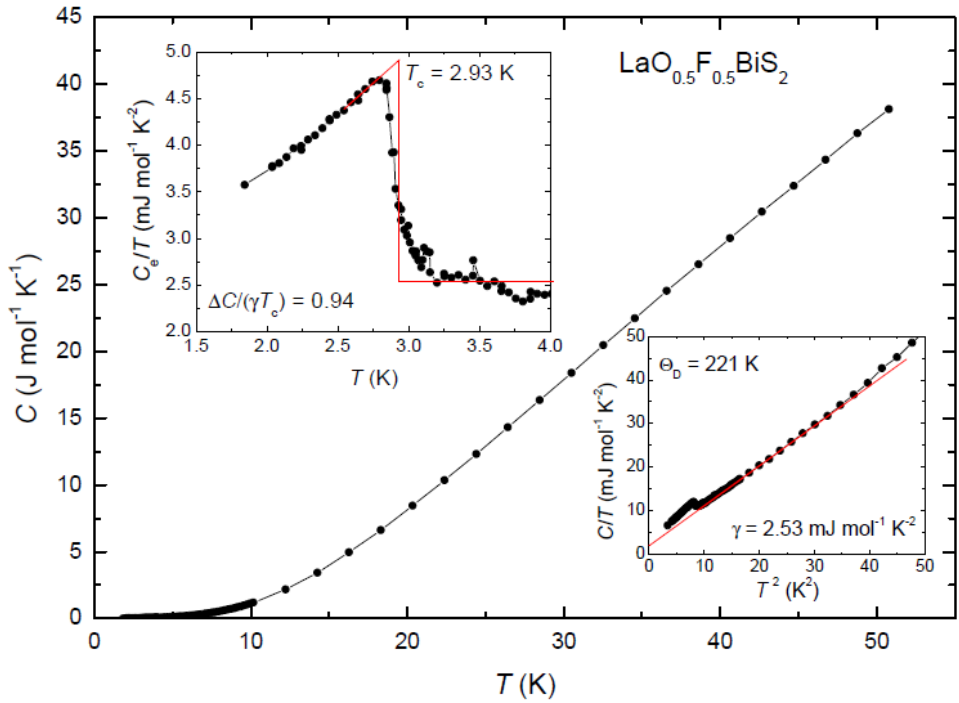

Figure 5. Specific heat $C$ vs temperature $T$ for $LaO_{0.5}F_{0.5}BiS_2$. A plot of $C/T$ vs $T^2$ is shown in the inset in the lower right hand part of the figure. The red line is a fit of the expression $C(T) = \gamma + \beta T^2$, which yields the values for $\gamma = 2.53$ mJ/mol-$K^2$ and $\Theta_D = 221$ K. A plot of $C_e$ vs $T$, where $C_e$ is the electronic contribution to the specific heat, in the vicinity of the superconducting transition is shown in the inset in the upper left hand side of the figure. An idealized entropy conserving construction yields $T_c = 2.93$ K and $\Delta C/(\gamma T_c) = 0.94$.

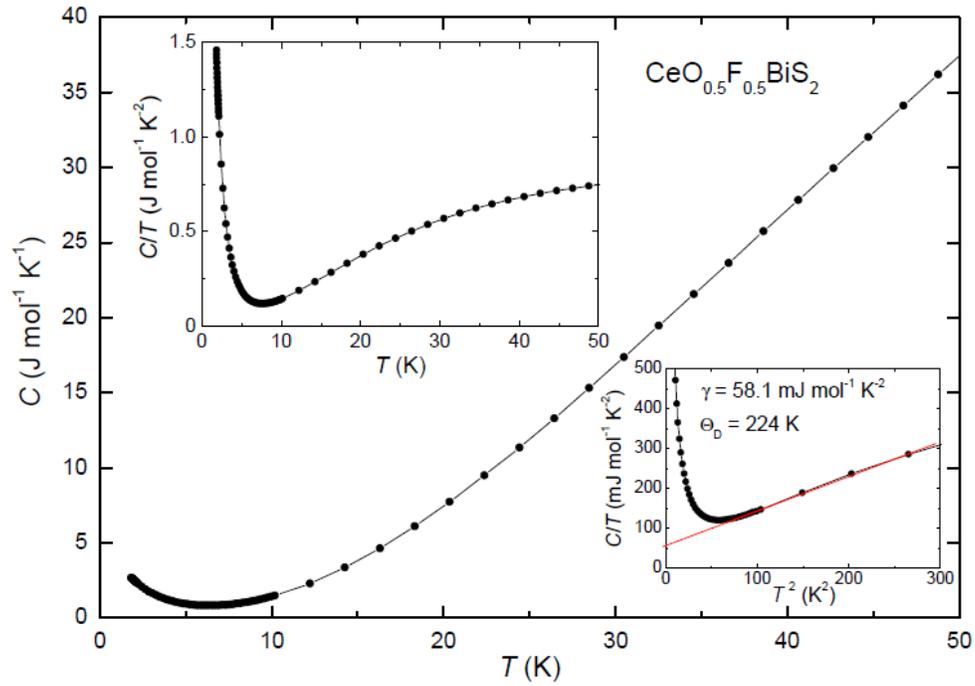

Figure 6. Specific heat $C$ vs temperature $T$ for CeO$_{0.5}$F$_{0.5}$BiS$_2$. A plot of $C/T$ vs $T^2$ is shown in the inset in the lower right hand part of the figure. The red line is a fit of the expression $C(T) = \gamma + \beta T^2$, which yields the values for $\gamma = 50.1$ mJ/mol-K$^2$ and $\Theta_D = 224$ K. A plot of $C$ vs $T$ is shown in the inset in the upper left hand side of the figure.

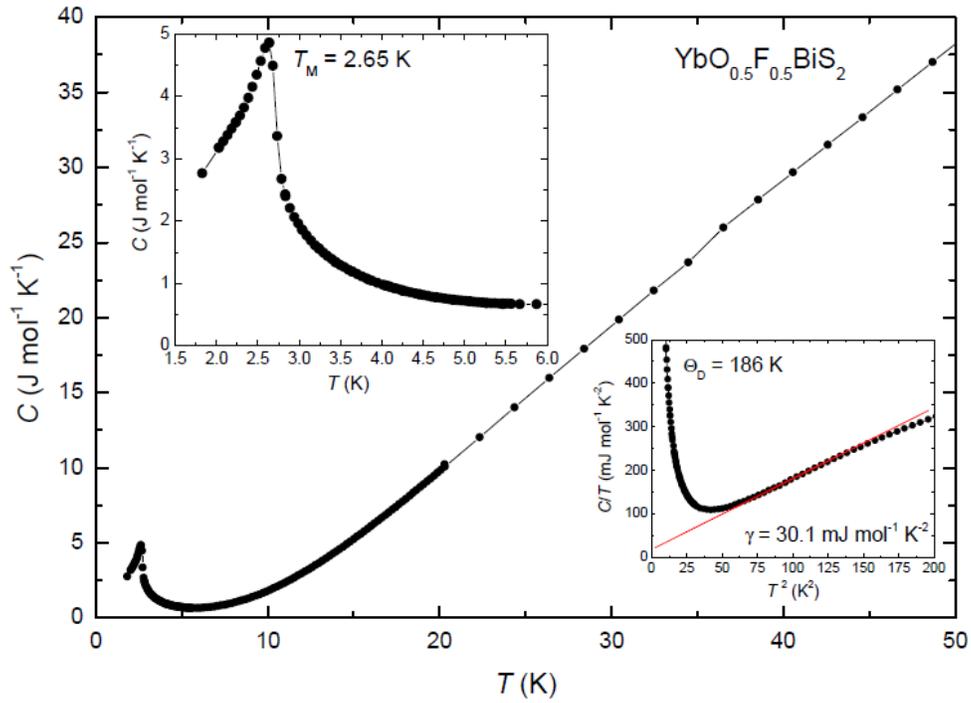

Figure 7. Specific heat $C$ vs temperature $T$ for YbO$_{0.5}$F$_{0.5}$BiS$_2$.  A plot of $C/T$ vs $\underline{T^2}$ is shown in the inset in the lower right hand part of the figure. The red line is a fit of the expression $C(T) = \gamma + \beta T^2$, which yields the values for $\gamma = 30.1$ mJ/mol-K$^2$ and $\Theta_D = 186$ K.  A plot of $C$ vs $T$, shown in the inset in the upper left hand side of the figure, reveals a sharp feature that is apparently associated with magnetic order that occurs below $T_M = 2.65$ K.